\documentclass[aps,prb,graphicx,twocolumn]{revtex4-1}
\pdfoutput=1
\usepackage[version=3]{mhchem} % Formula subscripts using \ce{}
\usepackage[T1]{fontenc}
\usepackage{siunitx}
\usepackage{graphicx}

\begin{document}

\author{Robin J. Dolleman}
\email{R.J.Dolleman@tudelft.nl}
\author{Santiago J. Cartamil-Bueno}
\author{Herre S. J. van der Zant}
\author{Peter G. Steeneken}
\affiliation{Kavli Institute of Nanoscience, Delft University of Technology, Lorentzweg 1, 2628CJ, Delft, The Netherlands}
\title{Graphene Gas Osmometers}

\begin{abstract}
Here it is shown that graphene membranes that separate 2 gases at identical pressure are deflected by osmotic pressure. The osmotic pressure is a consequence of differences in gas permeation rates into a graphene enclosed cavity. The deflection of the few layer graphene membranes is detected by an interferometric technique for measuring their tension-induced resonance frequency. Using a calibration measurement of the relation between resonance frequency and pressure, the time dependent osmotic pressure on the graphene is extracted. The osmotic pressure for different combinations of gases shows large differences that can be accounted for by a model based on the different gas permeation rates. Thus a graphene membrane based gas osmometer with a responsitivity of $\sim$60 kHz/mbar and nanoscale dimensions is demonstrated. 
\end{abstract}

\maketitle

\begin{figure}
\includegraphics{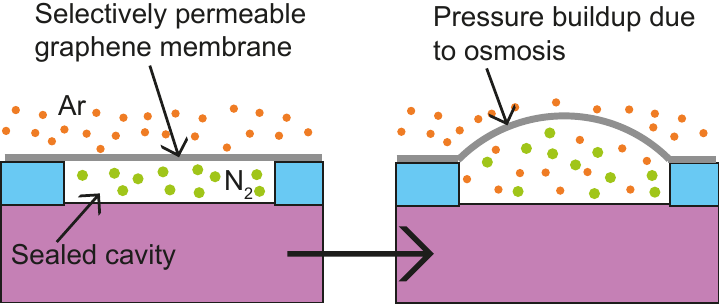}
\end{figure}

Graphene, a single layer of sp$^2$-bonded carbon atoms\cite{geim2007rise}, is impermeable to all gases \cite{bunch2008impermeable}. However, when pristine graphene is suspended over cavities in silicon dioxide, low-gas permeation rates between the cavity and the environment have been measured. Because the permeation rate was found to depend on the type of gas\cite{bunch2008impermeable,koenig2012selective,drahushuk2015analysis}, graphene enclosed cavities can therefore be selectively-permeable. When a selectively-permeable membrane separates different gases, osmotic gas flow causes a pressure difference across the membrane which is defined as the osmotic pressure \cite{kramer2012five}. The high Young's modulus\cite{lee2008measurement} and low bending rigidity cause a large pressure-induced frequency shift and deflection, which is beneficial for several types of pressure sensors\cite{zhu2013graphene,dolleman2015graphene,smith2013electromechanical}.

\begin{figure}
\includegraphics{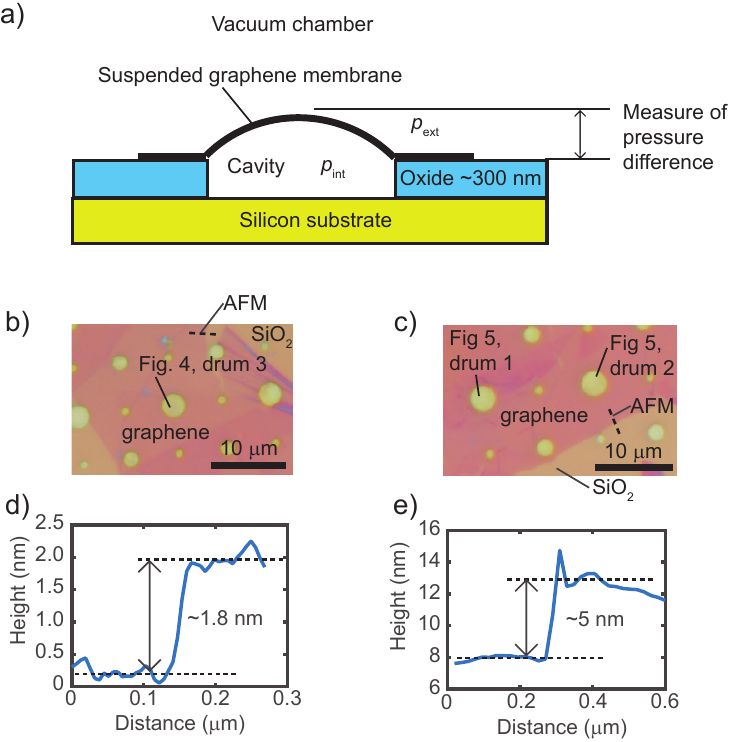}
\caption{a) Schematics of a graphene-based osmometer. b) Optical image of the graphene resonators presented in this study in Fig. \ref{fig:frequencies}. c) Optical image of the graphene resonators presented in Fig. \ref{fig:timeconstants}. d) Atomic force microscopy (AFM) trace corresponding to Fig. \ref{fig:fig1}b, showing that the graphene resonator has a thickness of about 1.8 nm. e) AFM trace corresponding to Fig. \ref{fig:fig1}c; these drums have a thickness of approximately 5 nm. \label{fig:fig1}}
\end{figure}

In this work we use graphene enclosed cavities to demonstrate osmotic pressure sensing for several combinations of gases, creating a nanoscale osmometer. For this purpose, graphene membranes are suspended over cavities etched in thermally grown silicon dioxide. A schematic device cross-section is shown in Fig. \ref{fig:fig1}a. A few-layer graphene flake with various thicknesses is exfoliated and transferred by a deterministic dry-transfer method\cite{castellanos2014deterministic} to enclose a cavity with a diameter of 3 $\mu$m (Fig. \ref{fig:fig1}b--e). It is found that such a device creates a selectively permeable system, without any further processing necessary. 

In the experiment, the gas outside the cavity is changed, while the pressure outside the cavity $p_{\mathrm{ext}}$ is kept constant. Deflections of the membrane due to external pressure changes are avoided and changes in the pressure difference $\Delta p=p_{\mathrm{int}}-p_{\mathrm{ext}}$  across the membrane can be solely attributed to changes in the internal pressure $p_{\mathrm{int}}$. Due to osmosis between the cavity and environment, it is observed that a pressure difference builds up over the membrane, even though the pressure on both sides is equal at the start of the experiment. 

\begin{figure}[t]
\includegraphics{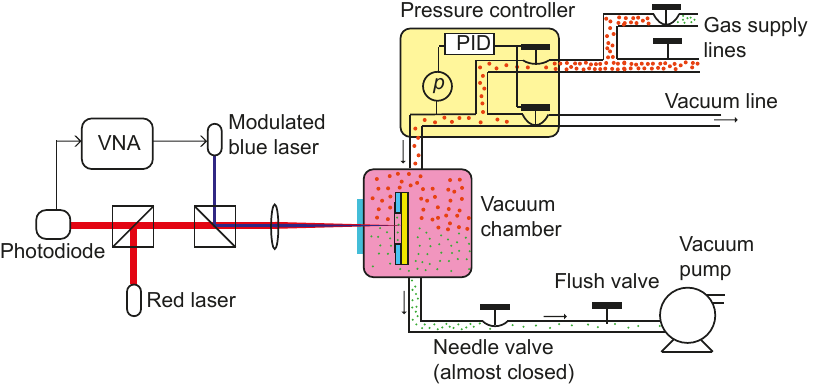}
\caption{Laser interferometer setup (left hand side of the figure) used to detect the resonance frequency and the vacuum chamber with the most important components for flushing the system at constant pressure (right hand side of the figure). \label{fig:figinter}}
\end{figure}
This pressure difference can be determined using the membrane's resonance frequency, that is measured by the interferometric measurement setup shown in Fig. \ref{fig:figinter}. A modulated blue laser provides opto-thermal actuation, while a red laser is used for interferometric readout of the deflection. A vector network analyzer (VNA) is used to measure the mechanical frequency response of the membrane\cite{castellanos2013single,davidovikj2016visualizing}. The sample is mounted in a vacuum chamber with optical access and a dual valve pressure controller is used to keep the pressure in the chamber ($p_{\mathrm{ext}}$) constant throughout the experiment. A vacuum line is connected to the chamber with a flush valve. The gas in the chamber is changed by switching the gas supply line and opening the flush valve. A needle valve restricts the flow to minimize pressure drops between the controller and the chamber. This prevents that the membrane deflects due to changes in $p_{\mathrm{ext}}$.

\begin{figure}[t]
\includegraphics{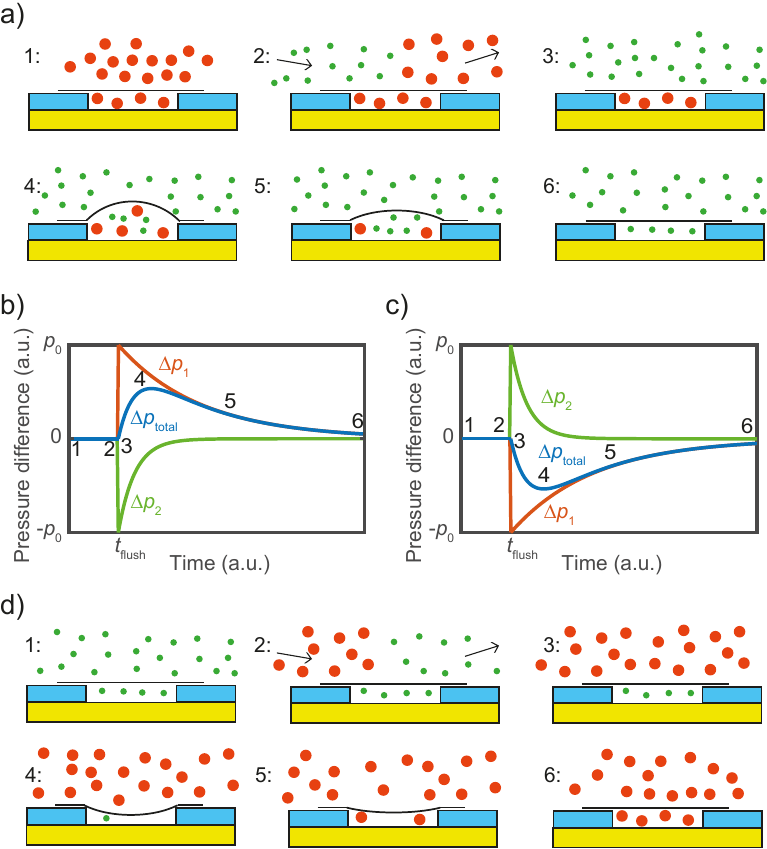}
\caption{\label{fig:fig2} a) Measurement sequence when replacing gas 1 (red) with a large permeation time constant $\tau_1$ by gas 2 (green) with a short permeation time constant $\tau_2$. b) Time dependent partial pressures differences $\Delta p_{\mathrm{part}}$ of both gases and the total osmotic pressure  $\Delta p$ as described by eq. (\ref{eq:pt}) for the measurement sequences depicted in Fig. \ref{fig:fig2}a. c) Partial pressures and total osmotic pressure for the sequence in Fig. \ref{fig:fig2}d. d) Measurement sequence when replacing gas 2 with a short permeation time constant $\tau_2$ by gas 1 (red) with a long permeation time constant $\tau_1$.}
\end{figure}

Fig. \ref{fig:fig2}a shows the measurement procedure for studying the time dependent osmotic pressure on the membrane. The sample is kept for a long time at a constant pressure in gas 1 (red), such that the internal and external pressure equalize $p_{\mathrm{ext}}=p_{\mathrm{int}}$ (Fig. \ref{fig:fig2}a1). The external gas 1 is replaced by gas 2 (green molecules) while keeping the pressure $p_{\mathrm{ext}}$ constant (Fig. \ref{fig:fig2}a2,3). This replacement is done rapidly to ensure that gas 1 remains present in the cavity at the same partial pressure as gas 2 in the vacuum chamber ($p_{1,\mathrm{int}} = p_{2,\mathrm{ext}}$).
If the permeation rate of gas 2 is higher than that of gas 1, gas 2 has a higher flux into the cavity than gas 1 flows out of it. Since the pressure inside the cavity is the sum of the partial pressures of gas 1 and 2, a positive pressure difference $\Delta p$ arises that is the osmotic pressure (Fig. \ref{fig:fig2}a4). Subsequently, gas 1 will leak out of the cavity at a slower rate (Fig. \ref{fig:fig2}a5) until gas 1 fully disappears and the pressure difference returns to zero $\Delta p \approx 0$ (Fig. \ref{fig:fig2}a6).
 
In a subsequent measurement gas 2 can be replaced by gas 1 in a similar manner which leads to the sequence shown in Fig. \ref{fig:fig2}d. The main difference is that in this case a negative pressure difference $\Delta p$ arises. Since permeation has exponential time dependence (see Supporting Information) the  pressure difference versus time $\Delta p(t)$ can be expressed by the partial pressure differences ($\Delta p_1$ and $\Delta p_2$) for each gas as function of time: $\Delta p_1 = p_0 \mathrm{e}^{-t/\tau_1}$ and $\Delta p_2 = - p_0 \mathrm{e}^{-t/\tau_2}$. Combining these equations gives for the total pressure difference:
\begin{equation} \label{eq:pt}
\Delta p(t) = \Delta p_1 + \Delta p_2 = p_0(\mathrm{e}^{-t /\tau_1} - \mathrm{e}^{- t/\tau_2})
\end{equation}
where $p_0$ is the constant pressure in the environment, $\tau_{1,2}$ are the leak-time constant inversely proportional to the permeability of gas 1 and gas 2, respectively. The expected time dependence of the osmotic pressure $\Delta p$ between two gases 1 and 2 with permeation rates $\tau_2$ and $\tau_1$ as described by Eq. (\ref{eq:pt}) is depicted in Fig. \ref{fig:fig2}b,c.

\begin{figure}[t]
\includegraphics{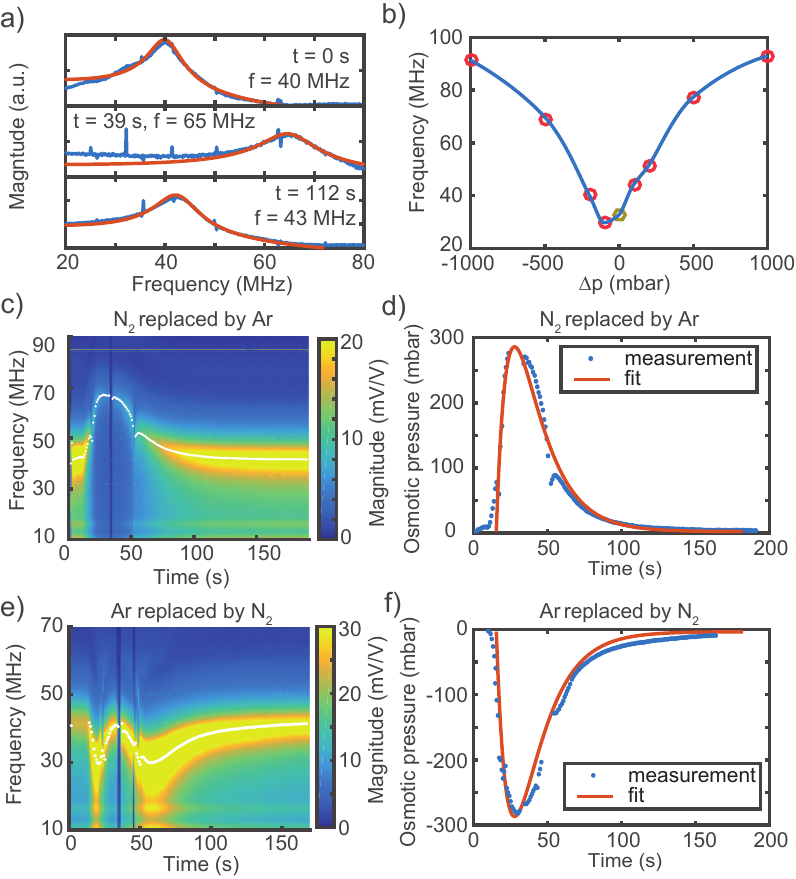}
\caption{Measurement of the osmotic pressure between argon and nitrogen for drum 3 (Fig. \ref{fig:fig1}b). a) Several frequency response functions from measurements (blue) and the fits (red) used to obtain the fundamental resonance frequency. b) Calibration curve used to convert the frequency into the pressure. c) Intensity plot of the frequency response function when nitrogen is replaced by argon in the chamber. White points show the extracted resonance frequency obtained from the fits. d) Osmotic pressure extracted from the experiment in Fig. \ref{fig:frequencies}c, fitted by a time-shifted version of eq. (\ref{eq:pt}). e) Intensity plot of the reverse experiment, where argon gas was replaced by nitrogen. f) Extracted osmotic pressure from the experiment in Fig. \ref{fig:frequencies}e. The fit from Fig. \ref{fig:frequencies}d is plotted with opposite sign. \label{fig:frequencies}}
\end{figure}

\begin{figure*}[t]
\includegraphics{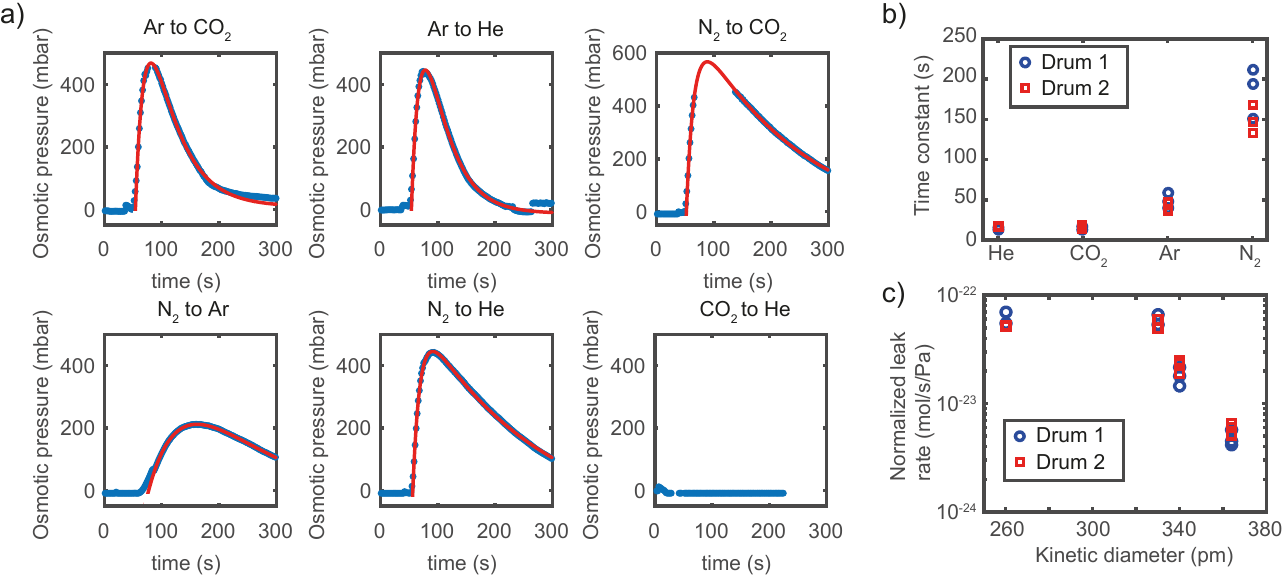}
\caption{a) Measurement sequences as in Fig. \ref{fig:fig2}a for 6 different gas combinations on a 5 nm thick drum (Drum 1) measured at 500 mbar. b Leak time constants $\tau$ extracted for 4 different gases using the fits in Fig. \ref{fig:timeconstants}a using two different 5 nm thick drums. Drum 2 is measured at 1000 mbar. An optical image of both drums is shown in Fig. \ref{fig:fig1}c. c) Normalized leak rates calculated from the leak time constants in Fig. \ref{fig:timeconstants}b. \label{fig:timeconstants}}
\end{figure*}

Figure \ref{fig:frequencies} shows the results of an experiment where nitrogen gas was replaced with argon gas and vice versa, at a constant chamber pressure of $p_{\mathrm{ext}} =  1000$ mbar. The resonance frequency is found by fitting the data to the frequency response function (Fig. \ref{fig:frequencies}a), which in turn yields the time-dependent resonance frequency. To extract the osmotic pressure from the experiment, the resonance-frequency versus pressure relation of the graphene membrane is calibrated by applying controlled gas pressure steps across the membrane (see Supporting Information). The resulting frequency-pressure difference relation is shown in Fig. \ref{fig:frequencies}b. Note, that the minimum in frequency does not correspond to a pressure difference of zero, but is shifted towards negative pressure differences and is around -100 mbar in this case. From the calibration curve, it is further concluded that this graphene-based osmometer has an average responsivity of approximately 60 kHz/mbar over the entire pressure range.

 Figure \ref{fig:frequencies}c,e shows the intensity plots of the frequency response function as function of time. White points on these figures indicate the resonance frequencies determined by the fits. The strong difference between these two curves is a consequence of the shifted calibration curve, which allows us to distinguish between positive and negative osmotic pressure. Therefore, from Fig. \ref{fig:frequencies}c we can conclude that argon was permeating into the cavity faster than nitrogen could escape, creating a positive pressure difference. In Fig. \ref{fig:frequencies}e, the frequency passes through a minimum twice; a clear indication that a negative pressure difference has formed over the membrane. In this case, argon was escaping the cavity faster than nitrogen could enter. From the time dependent fundamental resonance frequency and the calibration curve in Fig. \ref{fig:frequencies}b, the time dependent osmotic pressure can be extracted as shown in Fig. \ref{fig:frequencies}d,f.  Equation \ref{eq:pt} (adapted to include a time shift between the start of the measurement and the gas being replaced) is fitted against the data in Fig. \ref{fig:frequencies}d to extract the permeation time-constants of the gases \cite{sunqvist}: $\tau_{\mathrm{N_2}} = 19$ s and $\tau_{\mathrm{Ar}} = 8$ s. The osmotic pressure as function of time from this fit is plotted in Fig. \ref{fig:frequencies}f in good correspondence to the measurement result. This demonstrates that the osmotic pressure reverses sign when interchanging the gases in the experiment.

Figure \ref{fig:timeconstants}a shows experimental osmotic pressure versus time results for different gas combinations, extracted using the same method as in Fig. \ref{fig:frequencies} but on a different drum. Experiments were carried out with helium, argon, carbon dioxide and nitrogen gas. Equation \ref{eq:pt} is fitted to all the 6 osmotic pressure curves to extract the leak time constants as shown in Fig. \ref{fig:timeconstants}b. A factor of ~10 difference is observed in the permeation time constant of helium compared to that of nitrogen gas.  The full dataset with the frequency response functions for all combinations of gases is available in the Supporting Information.

It is important to note that the presented experiments cannot determine the exact leakage path of the gas molecules into the graphene cavity, although the results do allow to exclude some possible causes. If pores are present that are much larger than the molecular size, Graham's law for effusion predicts permeation rates proportional to the square root of the molecular mass ($\tau_1/\tau_2 \propto \sqrt{M_1/M_2}$). It is however observed, that carbon dioxide and helium have almost the same permeation rate, despite their large difference in molecular mass. On the other hand, carbon dioxide has a larger mass than nitrogen, but a lower permeation rate, again inconsistent with Graham's law for effusion. From this, we conclude that in this study permeation is not dominated by effusion through pores larger than the molecular size of the gases. For example, these kind of pores were studied by Celebi et. al\cite{celebi2014ultimate}, who found that Graham's law does hold in that case.

 Looking at the permeation rates for the different gases in Fig. \ref{fig:timeconstants}b, it is found that they follow the order of the kinetic diameters $d$ of the gases: ($d_{\rm He} = 260$ pm, $d_{\rm CO_2} = 330$ pm, $d_{\rm Ar} = 340$ pm, $d_{\rm N_2} = 364$ pm). Thus, gases with a kinetic diameter larger than $\sim$330 pm have a lower permeation rate than gases with a kinetic diameter smaller than $\sim$330 pm as shown in Fig. \ref{fig:timeconstants}c. This kind of selectivity in permeation is similar to the one observed by Koenig et. al. \cite{koenig2012selective,liu2013insights}, although the leak rates observed here are higher.

If the gas selectivity of the graphene enclosed cavities can be understood and engineered to a larger degree, for example by creating pores of controlled size \cite{koenig2012selective}, multiple semi-permeable membranes can be used for gas analysis. This can be achieved by filling these systems with a known gas and subsequently monitoring their time dependent osmotic pressure while exposing them to an unknown gas mixture. On the other hand, the gas concentration in the environment can be kept constant and the change in concentration in a very small volume can be determined. 

In conclusion, we have demonstrated osmotic pressure sensing with graphene enclosed cavities. The osmotic pressure is a consequence of differences in the permeation rate of the gases, resulting in a  spontaneous flux of gas against the pressure gradient. Due to the high flexibility and Young's modulus of graphene, the responsivity of the graphene osmometer is as high as 60 kHz/mbar. These types of graphene osmometers thus provide a route towards on-chip nanoscale gas analysis applications with high responsivity.

\section*{Acknowledgements}
We thank Dejan Davidovikj, Yaroslav Blanter and Farbod Alijani for discussions. The authors further thank the Dutch Technology Foundation (STW), which is part of the Netherlands Organisation for Scientific Research (NWO), and which is partly funded by the Ministry of Economic Affairs, for financially supporting this work. The research leading to these results has also received funding from the European Union Seventh Framework Programme under grant agreement number 604391 Graphene Flagship.
\bibliography{osmosis}

\begin{thebibliography}{15}%
\makeatletter
\providecommand \@ifxundefined [1]{%
 \@ifx{#1\undefined}
}%
\providecommand \@ifnum [1]{%
 \ifnum #1\expandafter \@firstoftwo
 \else \expandafter \@secondoftwo
 \fi
}%
\providecommand \@ifx [1]{%
 \ifx #1\expandafter \@firstoftwo
 \else \expandafter \@secondoftwo
 \fi
}%
\providecommand \natexlab [1]{#1}%
\providecommand \enquote  [1]{``#1''}%
\providecommand \bibnamefont  [1]{#1}%
\providecommand \bibfnamefont [1]{#1}%
\providecommand \citenamefont [1]{#1}%
\providecommand \href@noop [0]{\@secondoftwo}%
\providecommand \href [0]{\begingroup \@sanitize@url \@href}%
\providecommand \@href[1]{\@@startlink{#1}\@@href}%
\providecommand \@@href[1]{\endgroup#1\@@endlink}%
\providecommand \@sanitize@url [0]{\catcode `\\12\catcode `\$12\catcode
  `\&12\catcode `\#12\catcode `\^12\catcode `\_12\catcode `\%12\relax}%
\providecommand \@@startlink[1]{}%
\providecommand \@@endlink[0]{}%
\providecommand \url  [0]{\begingroup\@sanitize@url \@url }%
\providecommand \@url [1]{\endgroup\@href {#1}{\urlprefix }}%
\providecommand \urlprefix  [0]{URL }%
\providecommand \Eprint [0]{\href }%
\providecommand \doibase [0]{http://dx.doi.org/}%
\providecommand \selectlanguage [0]{\@gobble}%
\providecommand \bibinfo  [0]{\@secondoftwo}%
\providecommand \bibfield  [0]{\@secondoftwo}%
\providecommand \translation [1]{[#1]}%
\providecommand \BibitemOpen [0]{}%
\providecommand \bibitemStop [0]{}%
\providecommand \bibitemNoStop [0]{.\EOS\space}%
\providecommand \EOS [0]{\spacefactor3000\relax}%
\providecommand \BibitemShut  [1]{\csname bibitem#1\endcsname}%
\let\auto@bib@innerbib\@empty
%</preamble>
\bibitem [{\citenamefont {Geim}\ and\ \citenamefont
  {Novoselov}(2007)}]{geim2007rise}%
  \BibitemOpen
  \bibfield  {author} {\bibinfo {author} {\bibfnamefont {A.~K.}\ \bibnamefont
  {Geim}}\ and\ \bibinfo {author} {\bibfnamefont {K.~S.}\ \bibnamefont
  {Novoselov}},\ }\href@noop {} {\bibfield  {journal} {\bibinfo  {journal}
  {Nature materials}\ }\textbf {\bibinfo {volume} {6}},\ \bibinfo {pages} {183}
  (\bibinfo {year} {2007})}\BibitemShut {NoStop}%
\bibitem [{\citenamefont {Bunch}\ \emph {et~al.}(2008)\citenamefont {Bunch},
  \citenamefont {Verbridge}, \citenamefont {Alden}, \citenamefont {Van
  Der~Zande}, \citenamefont {Parpia}, \citenamefont {Craighead},\ and\
  \citenamefont {McEuen}}]{bunch2008impermeable}%
  \BibitemOpen
  \bibfield  {author} {\bibinfo {author} {\bibfnamefont {J.~S.}\ \bibnamefont
  {Bunch}}, \bibinfo {author} {\bibfnamefont {S.~S.}\ \bibnamefont
  {Verbridge}}, \bibinfo {author} {\bibfnamefont {J.~S.}\ \bibnamefont
  {Alden}}, \bibinfo {author} {\bibfnamefont {A.~M.}\ \bibnamefont {Van
  Der~Zande}}, \bibinfo {author} {\bibfnamefont {J.~M.}\ \bibnamefont
  {Parpia}}, \bibinfo {author} {\bibfnamefont {H.~G.}\ \bibnamefont
  {Craighead}}, \ and\ \bibinfo {author} {\bibfnamefont {P.~L.}\ \bibnamefont
  {McEuen}},\ }\href@noop {} {\bibfield  {journal} {\bibinfo  {journal} {Nano
  letters}\ }\textbf {\bibinfo {volume} {8}},\ \bibinfo {pages} {2458}
  (\bibinfo {year} {2008})}\BibitemShut {NoStop}%
\bibitem [{\citenamefont {Koenig}\ \emph {et~al.}(2012)\citenamefont {Koenig},
  \citenamefont {Wang}, \citenamefont {Pellegrino},\ and\ \citenamefont
  {Bunch}}]{koenig2012selective}%
  \BibitemOpen
  \bibfield  {author} {\bibinfo {author} {\bibfnamefont {S.~P.}\ \bibnamefont
  {Koenig}}, \bibinfo {author} {\bibfnamefont {L.}~\bibnamefont {Wang}},
  \bibinfo {author} {\bibfnamefont {J.}~\bibnamefont {Pellegrino}}, \ and\
  \bibinfo {author} {\bibfnamefont {J.~S.}\ \bibnamefont {Bunch}},\ }\href@noop
  {} {\bibfield  {journal} {\bibinfo  {journal} {Nature nanotechnology}\
  }\textbf {\bibinfo {volume} {7}},\ \bibinfo {pages} {728} (\bibinfo {year}
  {2012})}\BibitemShut {NoStop}%
\bibitem [{\citenamefont {Drahushuk}\ \emph {et~al.}(2015)\citenamefont
  {Drahushuk}, \citenamefont {Wang}, \citenamefont {Koenig}, \citenamefont
  {Bunch},\ and\ \citenamefont {Strano}}]{drahushuk2015analysis}%
  \BibitemOpen
  \bibfield  {author} {\bibinfo {author} {\bibfnamefont {L.~W.}\ \bibnamefont
  {Drahushuk}}, \bibinfo {author} {\bibfnamefont {L.}~\bibnamefont {Wang}},
  \bibinfo {author} {\bibfnamefont {S.~P.}\ \bibnamefont {Koenig}}, \bibinfo
  {author} {\bibfnamefont {J.~S.}\ \bibnamefont {Bunch}}, \ and\ \bibinfo
  {author} {\bibfnamefont {M.~S.}\ \bibnamefont {Strano}},\ }\href@noop {}
  {\bibfield  {journal} {\bibinfo  {journal} {ACS nano}\ } (\bibinfo {year}
  {2015})}\BibitemShut {NoStop}%
\bibitem [{\citenamefont {Kramer}\ and\ \citenamefont
  {Myers}(2012)}]{kramer2012five}%
  \BibitemOpen
  \bibfield  {author} {\bibinfo {author} {\bibfnamefont {E.~M.}\ \bibnamefont
  {Kramer}}\ and\ \bibinfo {author} {\bibfnamefont {D.~R.}\ \bibnamefont
  {Myers}},\ }\href@noop {} {\bibfield  {journal} {\bibinfo  {journal}
  {American Journal of Physics}\ }\textbf {\bibinfo {volume} {80}},\ \bibinfo
  {pages} {694} (\bibinfo {year} {2012})}\BibitemShut {NoStop}%
\bibitem [{\citenamefont {Lee}\ \emph {et~al.}(2008)\citenamefont {Lee},
  \citenamefont {Wei}, \citenamefont {Kysar},\ and\ \citenamefont
  {Hone}}]{lee2008measurement}%
  \BibitemOpen
  \bibfield  {author} {\bibinfo {author} {\bibfnamefont {C.}~\bibnamefont
  {Lee}}, \bibinfo {author} {\bibfnamefont {X.}~\bibnamefont {Wei}}, \bibinfo
  {author} {\bibfnamefont {J.~W.}\ \bibnamefont {Kysar}}, \ and\ \bibinfo
  {author} {\bibfnamefont {J.}~\bibnamefont {Hone}},\ }\href@noop {} {\bibfield
   {journal} {\bibinfo  {journal} {science}\ }\textbf {\bibinfo {volume}
  {321}},\ \bibinfo {pages} {385} (\bibinfo {year} {2008})}\BibitemShut
  {NoStop}%
\bibitem [{\citenamefont {Zhu}\ \emph {et~al.}(2013)\citenamefont {Zhu},
  \citenamefont {Ghatkesar}, \citenamefont {Zhang},\ and\ \citenamefont
  {Janssen}}]{zhu2013graphene}%
  \BibitemOpen
  \bibfield  {author} {\bibinfo {author} {\bibfnamefont {S.-E.}\ \bibnamefont
  {Zhu}}, \bibinfo {author} {\bibfnamefont {M.~K.}\ \bibnamefont {Ghatkesar}},
  \bibinfo {author} {\bibfnamefont {C.}~\bibnamefont {Zhang}}, \ and\ \bibinfo
  {author} {\bibfnamefont {G.}~\bibnamefont {Janssen}},\ }\href@noop {}
  {\bibfield  {journal} {\bibinfo  {journal} {Applied Physics Letters}\
  }\textbf {\bibinfo {volume} {102}},\ \bibinfo {pages} {161904} (\bibinfo
  {year} {2013})}\BibitemShut {NoStop}%
\bibitem [{\citenamefont {Dolleman}\ \emph {et~al.}(2016)\citenamefont
  {Dolleman}, \citenamefont {Davidovikj}, \citenamefont {Cartamil-Bueno},
  \citenamefont {van~der Zant},\ and\ \citenamefont
  {Steeneken}}]{dolleman2015graphene}%
  \BibitemOpen
  \bibfield  {author} {\bibinfo {author} {\bibfnamefont {R.~J.}\ \bibnamefont
  {Dolleman}}, \bibinfo {author} {\bibfnamefont {D.}~\bibnamefont
  {Davidovikj}}, \bibinfo {author} {\bibfnamefont {S.~J.}\ \bibnamefont
  {Cartamil-Bueno}}, \bibinfo {author} {\bibfnamefont {H.~S.}\ \bibnamefont
  {van~der Zant}}, \ and\ \bibinfo {author} {\bibfnamefont {P.~G.}\
  \bibnamefont {Steeneken}},\ }\href@noop {} {\bibfield  {journal} {\bibinfo
  {journal} {Nano letters}\ }\textbf {\bibinfo {volume} {16}},\ \bibinfo
  {pages} {568} (\bibinfo {year} {2016})}\BibitemShut {NoStop}%
\bibitem [{\citenamefont {Smith}\ \emph {et~al.}(2013)\citenamefont {Smith},
  \citenamefont {Niklaus}, \citenamefont {Paussa}, \citenamefont {Vaziri},
  \citenamefont {Fischer}, \citenamefont {Sterner}, \citenamefont {Forsberg},
  \citenamefont {Delin}, \citenamefont {Esseni}, \citenamefont {Palestri},
  \citenamefont {Ostling},\ and\ \citenamefont
  {Lemme}}]{smith2013electromechanical}%
  \BibitemOpen
  \bibfield  {author} {\bibinfo {author} {\bibfnamefont {A.}~\bibnamefont
  {Smith}}, \bibinfo {author} {\bibfnamefont {F.}~\bibnamefont {Niklaus}},
  \bibinfo {author} {\bibfnamefont {A.}~\bibnamefont {Paussa}}, \bibinfo
  {author} {\bibfnamefont {S.}~\bibnamefont {Vaziri}}, \bibinfo {author}
  {\bibfnamefont {A.~C.}\ \bibnamefont {Fischer}}, \bibinfo {author}
  {\bibfnamefont {M.}~\bibnamefont {Sterner}}, \bibinfo {author} {\bibfnamefont
  {F.}~\bibnamefont {Forsberg}}, \bibinfo {author} {\bibfnamefont
  {A.}~\bibnamefont {Delin}}, \bibinfo {author} {\bibfnamefont
  {D.}~\bibnamefont {Esseni}}, \bibinfo {author} {\bibfnamefont
  {P.}~\bibnamefont {Palestri}}, \bibinfo {author} {\bibfnamefont
  {M.}~\bibnamefont {Ostling}}, \ and\ \bibinfo {author} {\bibfnamefont
  {M.}~\bibnamefont {Lemme}},\ }\href@noop {} {\bibfield  {journal} {\bibinfo
  {journal} {Nano letters}\ }\textbf {\bibinfo {volume} {13}},\ \bibinfo
  {pages} {3237} (\bibinfo {year} {2013})}\BibitemShut {NoStop}%
\bibitem [{\citenamefont {Castellanos-Gomez}\ \emph {et~al.}(2014)\citenamefont
  {Castellanos-Gomez}, \citenamefont {Buscema}, \citenamefont {Molenaar},
  \citenamefont {Singh}, \citenamefont {Janssen}, \citenamefont {van~der
  Zant},\ and\ \citenamefont {Steele}}]{castellanos2014deterministic}%
  \BibitemOpen
  \bibfield  {author} {\bibinfo {author} {\bibfnamefont {A.}~\bibnamefont
  {Castellanos-Gomez}}, \bibinfo {author} {\bibfnamefont {M.}~\bibnamefont
  {Buscema}}, \bibinfo {author} {\bibfnamefont {R.}~\bibnamefont {Molenaar}},
  \bibinfo {author} {\bibfnamefont {V.}~\bibnamefont {Singh}}, \bibinfo
  {author} {\bibfnamefont {L.}~\bibnamefont {Janssen}}, \bibinfo {author}
  {\bibfnamefont {H.~S.}\ \bibnamefont {van~der Zant}}, \ and\ \bibinfo
  {author} {\bibfnamefont {G.~A.}\ \bibnamefont {Steele}},\ }\href@noop {}
  {\bibfield  {journal} {\bibinfo  {journal} {2D Materials}\ }\textbf {\bibinfo
  {volume} {1}},\ \bibinfo {pages} {011002} (\bibinfo {year}
  {2014})}\BibitemShut {NoStop}%
\bibitem [{\citenamefont {Castellanos-Gomez}\ \emph {et~al.}(2013)\citenamefont
  {Castellanos-Gomez}, \citenamefont {van Leeuwen}, \citenamefont {Buscema},
  \citenamefont {van~der Zant}, \citenamefont {Steele},\ and\ \citenamefont
  {Venstra}}]{castellanos2013single}%
  \BibitemOpen
  \bibfield  {author} {\bibinfo {author} {\bibfnamefont {A.}~\bibnamefont
  {Castellanos-Gomez}}, \bibinfo {author} {\bibfnamefont {R.}~\bibnamefont {van
  Leeuwen}}, \bibinfo {author} {\bibfnamefont {M.}~\bibnamefont {Buscema}},
  \bibinfo {author} {\bibfnamefont {H.~S.}\ \bibnamefont {van~der Zant}},
  \bibinfo {author} {\bibfnamefont {G.~A.}\ \bibnamefont {Steele}}, \ and\
  \bibinfo {author} {\bibfnamefont {W.~J.}\ \bibnamefont {Venstra}},\
  }\href@noop {} {\bibfield  {journal} {\bibinfo  {journal} {Advanced
  Materials}\ }\textbf {\bibinfo {volume} {25}},\ \bibinfo {pages} {6719}
  (\bibinfo {year} {2013})}\BibitemShut {NoStop}%
\bibitem [{\citenamefont {Davidovikj}\ \emph {et~al.}(2016)\citenamefont
  {Davidovikj}, \citenamefont {Slim}, \citenamefont {Cartamil-Bueno},
  \citenamefont {van~der Zant}, \citenamefont {Steeneken},\ and\ \citenamefont
  {Venstra}}]{davidovikj2016visualizing}%
  \BibitemOpen
  \bibfield  {author} {\bibinfo {author} {\bibfnamefont {D.}~\bibnamefont
  {Davidovikj}}, \bibinfo {author} {\bibfnamefont {J.~J.}\ \bibnamefont
  {Slim}}, \bibinfo {author} {\bibfnamefont {S.~J.}\ \bibnamefont
  {Cartamil-Bueno}}, \bibinfo {author} {\bibfnamefont {H.~S.}\ \bibnamefont
  {van~der Zant}}, \bibinfo {author} {\bibfnamefont {P.~G.}\ \bibnamefont
  {Steeneken}}, \ and\ \bibinfo {author} {\bibfnamefont {W.~J.}\ \bibnamefont
  {Venstra}},\ }\href@noop {} {\bibfield  {journal} {\bibinfo  {journal} {Nano
  letters}\ }\textbf {\bibinfo {volume} {16}},\ \bibinfo {pages} {2768}
  (\bibinfo {year} {2016})}\BibitemShut {NoStop}%
\bibitem [{\citenamefont {Sundqvist}()}]{sunqvist}%
  \BibitemOpen
  \bibfield  {author} {\bibinfo {author} {\bibfnamefont {P.}~\bibnamefont
  {Sundqvist}},\ }\href@noop {} {\enquote {\bibinfo {title} {Exponential
  curve-fitting, without start-guess},}\ }\bibinfo {note} {Retrieved 14 March
  2016 from:
  \url{http://www.mathworks.com/matlabcentral/fileexchange/21959-exponential-fit--without-start-guess}}\BibitemShut
  {NoStop}%
\bibitem [{\citenamefont {Celebi}\ \emph {et~al.}(2014)\citenamefont {Celebi},
  \citenamefont {Buchheim}, \citenamefont {Wyss}, \citenamefont {Droudian},
  \citenamefont {Gasser}, \citenamefont {Shorubalko}, \citenamefont {Kye},
  \citenamefont {Lee},\ and\ \citenamefont {Park}}]{celebi2014ultimate}%
  \BibitemOpen
  \bibfield  {author} {\bibinfo {author} {\bibfnamefont {K.}~\bibnamefont
  {Celebi}}, \bibinfo {author} {\bibfnamefont {J.}~\bibnamefont {Buchheim}},
  \bibinfo {author} {\bibfnamefont {R.~M.}\ \bibnamefont {Wyss}}, \bibinfo
  {author} {\bibfnamefont {A.}~\bibnamefont {Droudian}}, \bibinfo {author}
  {\bibfnamefont {P.}~\bibnamefont {Gasser}}, \bibinfo {author} {\bibfnamefont
  {I.}~\bibnamefont {Shorubalko}}, \bibinfo {author} {\bibfnamefont {J.-I.}\
  \bibnamefont {Kye}}, \bibinfo {author} {\bibfnamefont {C.}~\bibnamefont
  {Lee}}, \ and\ \bibinfo {author} {\bibfnamefont {H.~G.}\ \bibnamefont
  {Park}},\ }\href@noop {} {\bibfield  {journal} {\bibinfo  {journal}
  {Science}\ }\textbf {\bibinfo {volume} {344}},\ \bibinfo {pages} {289}
  (\bibinfo {year} {2014})}\BibitemShut {NoStop}%
\bibitem [{\citenamefont {Liu}\ \emph {et~al.}(2013)\citenamefont {Liu},
  \citenamefont {Dai},\ and\ \citenamefont {Jiang}}]{liu2013insights}%
  \BibitemOpen
  \bibfield  {author} {\bibinfo {author} {\bibfnamefont {H.}~\bibnamefont
  {Liu}}, \bibinfo {author} {\bibfnamefont {S.}~\bibnamefont {Dai}}, \ and\
  \bibinfo {author} {\bibfnamefont {D.}~\bibnamefont {Jiang}},\ }\href@noop {}
  {\bibfield  {journal} {\bibinfo  {journal} {Nanoscale}\ }\textbf {\bibinfo
  {volume} {5}},\ \bibinfo {pages} {9984} (\bibinfo {year} {2013})}\BibitemShut
  {NoStop}%
\end{thebibliography}%


\providecommand*\mcitethebibliography{\thebibliography}
\csname @ifundefined\endcsname{endmcitethebibliography}
  {\let\endmcitethebibliography\endthebibliography}{}
\begin{mcitethebibliography}{2}
\providecommand*\natexlab[1]{#1}
\providecommand*\mciteSetBstSublistMode[1]{}
\providecommand*\mciteSetBstMaxWidthForm[2]{}
\providecommand*\mciteBstWouldAddEndPuncttrue
  {\def\EndOfBibitem{\unskip.}}
\providecommand*\mciteBstWouldAddEndPunctfalse
  {\let\EndOfBibitem\relax}
\providecommand*\mciteSetBstMidEndSepPunct[3]{}
\providecommand*\mciteSetBstSublistLabelBeginEnd[3]{}
\providecommand*\EndOfBibitem{}
\mciteSetBstSublistMode{f}
\mciteSetBstMaxWidthForm{subitem}{(\alph{mcitesubitemcount})}
\mciteSetBstSublistLabelBeginEnd
  {\mcitemaxwidthsubitemform\space}
  {\relax}
  {\relax}

\bibitem[Dolleman et~al.(2016)Dolleman, Davidovikj, Cartamil-Bueno, van~der
  Zant, and Steeneken]{dolleman2015graphene}
Dolleman,~R.~J.; Davidovikj,~D.; Cartamil-Bueno,~S.~J.; van~der Zant,~H.~S.;
  Steeneken,~P.~G. \emph{Nano letters} \textbf{2016}, \emph{16}, 568--571\relax
\mciteBstWouldAddEndPuncttrue
\mciteSetBstMidEndSepPunct{\mcitedefaultmidpunct}
{\mcitedefaultendpunct}{\mcitedefaultseppunct}\relax
\EndOfBibitem
\end{mcitethebibliography}


\begin{thebibliography}{0}%
\makeatletter
\providecommand \@ifxundefined [1]{%
 \@ifx{#1\undefined}
}%
\providecommand \@ifnum [1]{%
 \ifnum #1\expandafter \@firstoftwo
 \else \expandafter \@secondoftwo
 \fi
}%
\providecommand \@ifx [1]{%
 \ifx #1\expandafter \@firstoftwo
 \else \expandafter \@secondoftwo
 \fi
}%
\providecommand \natexlab [1]{#1}%
\providecommand \enquote  [1]{``#1''}%
\providecommand \bibnamefont  [1]{#1}%
\providecommand \bibfnamefont [1]{#1}%
\providecommand \citenamefont [1]{#1}%
\providecommand \href@noop [0]{\@secondoftwo}%
\providecommand \href [0]{\begingroup \@sanitize@url \@href}%
\providecommand \@href[1]{\@@startlink{#1}\@@href}%
\providecommand \@@href[1]{\endgroup#1\@@endlink}%
\providecommand \@sanitize@url [0]{\catcode `\\12\catcode `\$12\catcode
  `\&12\catcode `\#12\catcode `\^12\catcode `\_12\catcode `\%12\relax}%
\providecommand \@@startlink[1]{}%
\providecommand \@@endlink[0]{}%
\providecommand \url  [0]{\begingroup\@sanitize@url \@url }%
\providecommand \@url [1]{\endgroup\@href {#1}{\urlprefix }}%
\providecommand \urlprefix  [0]{URL }%
\providecommand \Eprint [0]{\href }%
\providecommand \doibase [0]{http://dx.doi.org/}%
\providecommand \selectlanguage [0]{\@gobble}%
\providecommand \bibinfo  [0]{\@secondoftwo}%
\providecommand \bibfield  [0]{\@secondoftwo}%
\providecommand \translation [1]{[#1]}%
\providecommand \BibitemOpen [0]{}%
\providecommand \bibitemStop [0]{}%
\providecommand \bibitemNoStop [0]{.\EOS\space}%
\providecommand \EOS [0]{\spacefactor3000\relax}%
\providecommand \BibitemShut  [1]{\csname bibitem#1\endcsname}%
\let\auto@bib@innerbib\@empty
%</preamble>
\end{thebibliography}%



%\section*{Associated Content}
%Supporting Information Available: Derivation of Eq. \ref{eq:pt}, explanation of the calibration procedure used to obtain Fig. \ref{fig:frequencies}b, a discussion on the squeeze-film effect and the full datasets containing the frequency response functions corresponding to Figs. \ref{fig:frequencies}--\ref{fig:timeconstants}. This material is available free of charge via the Internet at http://pubs.acs.org.
\onecolumngrid
\pagebreak

\setcounter{equation}{0}
\setcounter{figure}{0}
\setcounter{table}{0}
\makeatletter
\renewcommand{\theequation}{S\arabic{equation}}
\renewcommand{\thefigure}{S\arabic{figure}}

\section*{Supporting Information: Graphene Gas Osmometers}

In this text equation 1 is derived, which decribes the pressure difference as a function of time over the membrane. Then the calibration procedure is explained which is used to extract the relation between pressure difference and resonance frequency. Finally, full raw datasets are presented for all the combinations of gases shown in Fig. 4 and 5 in the main text.

\section*{S1. Derivation of equation 1}
Dalton's law states that for a mixture of gases the total pressure $p_{\mathrm{tot}}$ is equal to the sum of the partial pressures of the individual components:
\begin{equation}\label{eq:ptot}
p_{\mathrm{tot}} = p_1 + p_2,
\end{equation}
where $p_1$ is the partial pressure of gas 1 and $p_2$ the partial pressure of gas 2. This allows us to calculate the pressure inside the cavity. If solubility of the material can be ignored, permeation through a membrane can be described by:
\begin{equation}
 \frac{\mathrm{d} n}{\mathrm{d} t} = - \mathcal{P} \Delta p A
 \end{equation}
where $ \frac{\mathrm{d} n}{\mathrm{d} t} $ is the flux through the membrane, $\Delta p$ the pressure difference over the membrane, $\mathcal{P}$ the permeability and $A$ the surface area of the membrane.

If it is assumed that the mixture of gases is ideal, according to eq. \ref{eq:ptot} we can also describe the permeation by the partial pressure differences between the cavity and surroundings. For this the following first order equations apply:
\begin{equation}\label{eq:per1}
\frac{\mathrm{d} \Delta p_1}{\mathrm{d} t} = -\frac{1}{\tau_1} \Delta p_1,
\end{equation}
\begin{equation}\label{eq:per2}
\frac{\mathrm{d} \Delta p_2}{\mathrm{d} t} = -\frac{1}{\tau_2} \Delta p_2,
\end{equation}
for which we define the leak time constants $\tau$:
\begin{equation}
 \tau = \frac{\mathcal{P} A RT}{V},
\end{equation}
where $R$ is the gas constant, $T$ the temperature and $V$ the volume of the cavity. These equations have the following solutions:
\begin{equation}
\Delta p_1(t)= k_1 \mathrm{e}^{-t/\tau_1} = p_{1,\mathrm{int}} - p_{1,\mathrm{ext}},
\end{equation}
\begin{equation}
\Delta p_2(t) = k_2 \mathrm{e}^{-t/\tau_2} = p_{2,\mathrm{int}} - p_{2,\mathrm{ext}},.
\end{equation}
Now we assume the initial conditions of the experiment. At $t=0$ the cavity is filled with pure gas 1 and the outside with pure gas 2 at pressure $p_0$. Substituting $t=0$ into both solutions gives that $\Delta p_1(0) = k_1 = p_0$ and $\Delta p_2(0) = k_2 = -p_0$. Adding the partial pressure differences over the membrane according to eq. \ref{eq:ptot} gives the total pressure difference as function of time:
\begin{equation}
\Delta p (t) = p_0 \mathrm{e}^{-t/\tau_1} - p_0 \mathrm{e}^{-t/\tau_2} = p_0 (\mathrm{e}^{-t/\tau_1} - \mathrm{e}^{-t/\tau_2} )
\end{equation}

\begin{figure}[h!]
\includegraphics{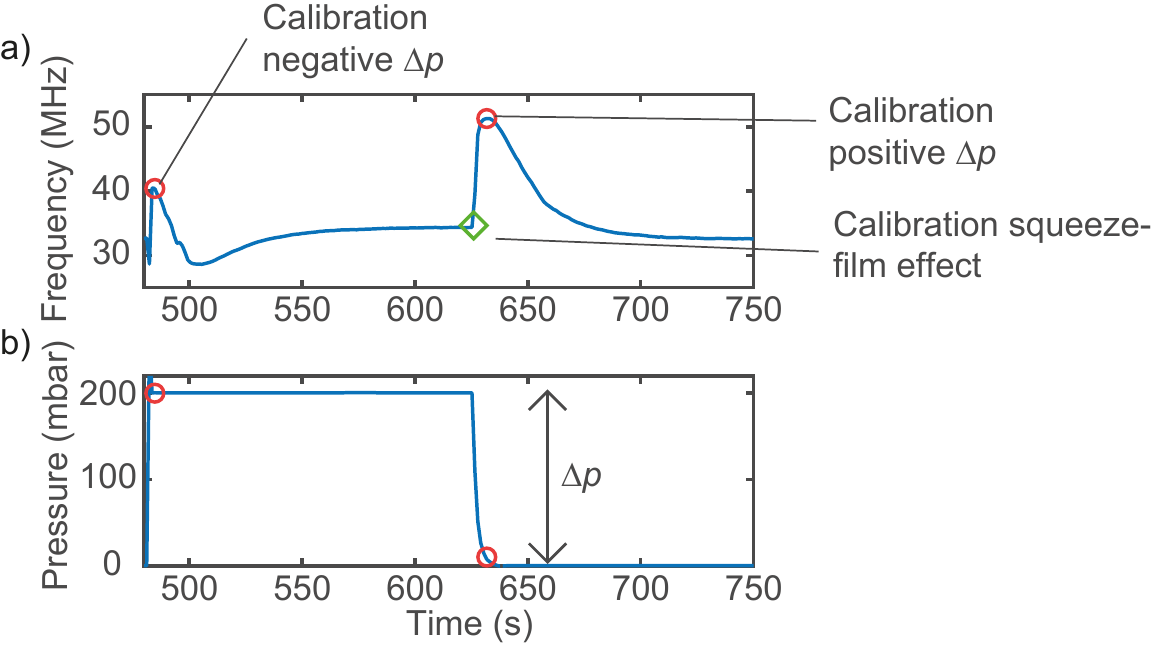}
\caption{Calibration procedure for obtaining the frequency--pressure difference relation in the case of 200 mbar steps. a) Resonance frequency as function of time, when 2 pressure steps are applied. b) Pressure signal measured in the chamber. The results of this procedure is the frequency--pressure difference relation as shown in Fig. 4b in the main text. \label{fig:cal}}
\end{figure}

\section*{S2. Calibration procedure}
In order to arrive at the calibration curve of Fig. 4b in the main text, the following procedure is used. The membrane is kept under vacuum long enough for any remaining gas to permeate out of the cavity. Then, the pressure in the environment is stepped up, after which the immediate change in resonance frequency is used as calibration point. The pressure difference over the membrane is negative in this case. After the permeation has caused the pressure difference over the membrane to become zero, the pressure is rapidly reduced below 1 mbar. This gives a calibration point for positive pressure difference.  The minimum in frequency does not correspond to a pressure difference of zero. This is either a result from the mechanical properties of the membrane or remaining gas inside the cavity. Regardless of the cause, the shifted curve is useful to determine whether the pressure difference during the experiment was positive or negative. 

For drum 3 (Fig. 4 in the main text), a shift in the calibration curve is observed towards negative pressure differences.  The cause of this shift is unknown and its value drifts slowly as a function of time. Since the calibration was performed 50 minutes after the measurements shown in this figure, it is necessary to correct for this effect. Therefore, the calibration curve was corrected by an additional -78 mbar, thereby ensuring that the osmotic pressure difference is zero at the start of both experiments. Drum 1 and 2 (Fig. 5 in the main text) show a similar shift of the frequency minimum, but the drift was considerably smaller. Therefore no correction of the calibation curve was necessary for these drums.

\section*{S3. Squeeze film effect}
\begin{figure*}[h!]
\includegraphics{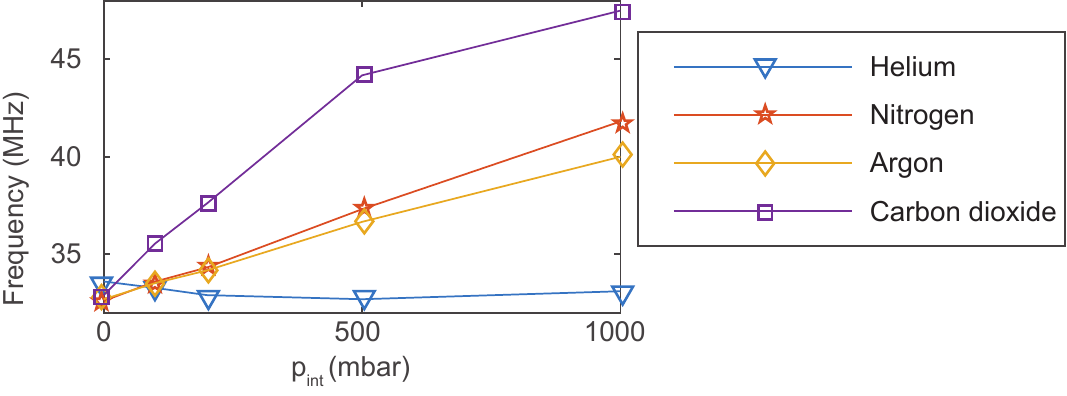}
\caption{Gas and pressure dependence of the frequency at a pressure difference of zero for drum 3. \label{fig:sqz}}
\end{figure*}

Due to the squeeze film effect, it is expected that the resonance frequency is also a function of the pressure inside the cavity \cite{dolleman2015graphene}. To examine this effect, the calibration procedure can be used. When a certain pressure step is applied, we wait long enough for the pressure difference to become close to zero (Fig. \ref{fig:cal}). This gives the calibration curve for the squeeze film effect as shown in Fig. \ref{fig:sqz} for drum 3. These curves show that the frequency shifts can not fully be attributed to pressure differences $\Delta p$ that induces tenstion to the membrane, but are also partly caused by the squeeze film effect that only depends on $p_{\mathrm{int}}$. 

\section*{S4. Full raw dataset of drum 3 in the main text}
\begin{figure*}[h!]
\includegraphics{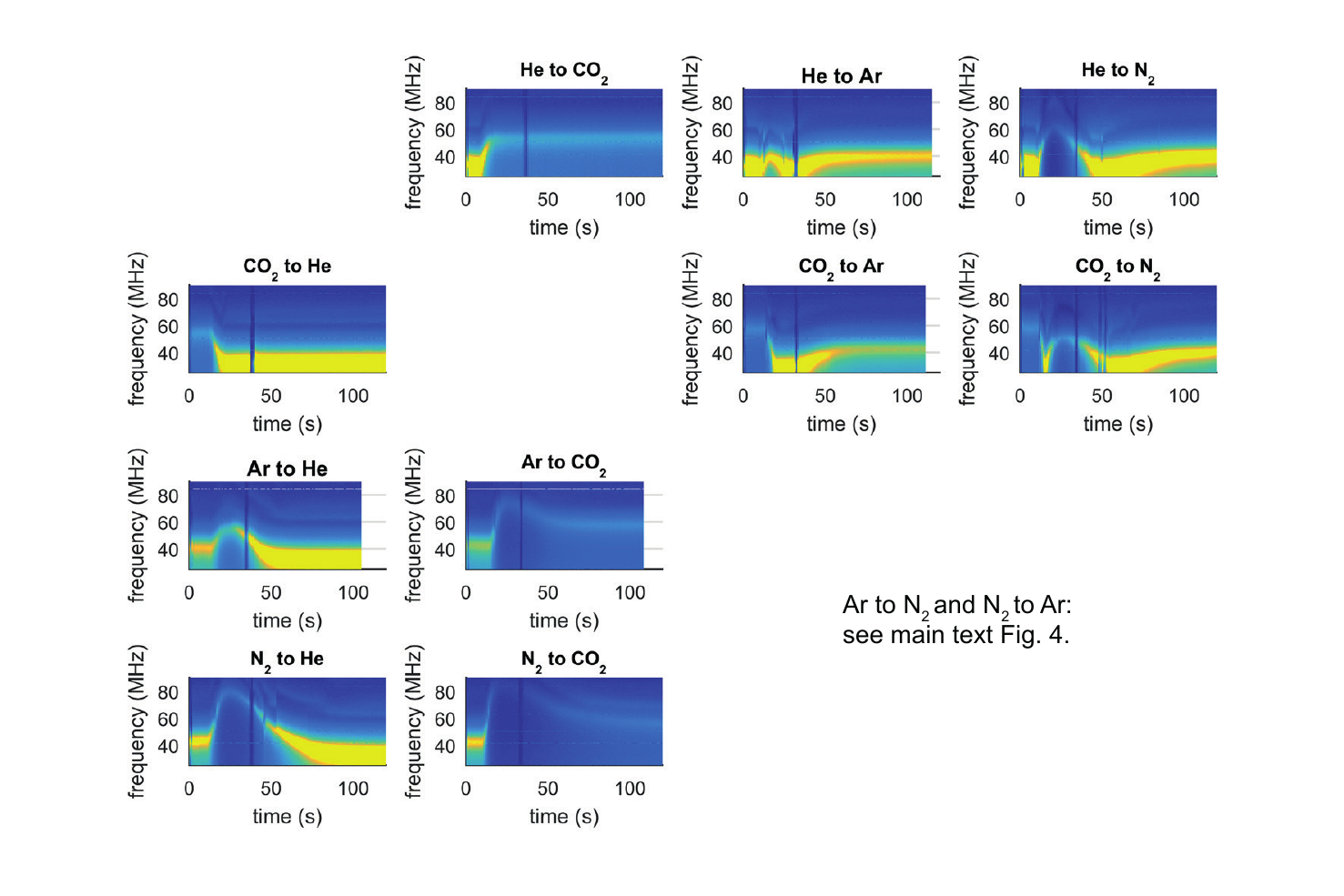}
\caption{Full raw dataset corresponding to the drum in Figure 4 in the main text, measurements between argon and nitrogen are presented in the main text.\label{fig:thindrum}}
\end{figure*}

Figure \ref{fig:thindrum} shows the full dataset of the drum presented in Fig. 4 in the main text.  This was measured using the calibration procedure, by taking the frequency when the membrane is fully relaxed ($\Delta p = 0$). This gives the result shown in Fig. \ref{fig:sqz}.

\section*{S5. Full raw datasets of drum 1 and 2 in the main text}
\begin{figure*}
\includegraphics{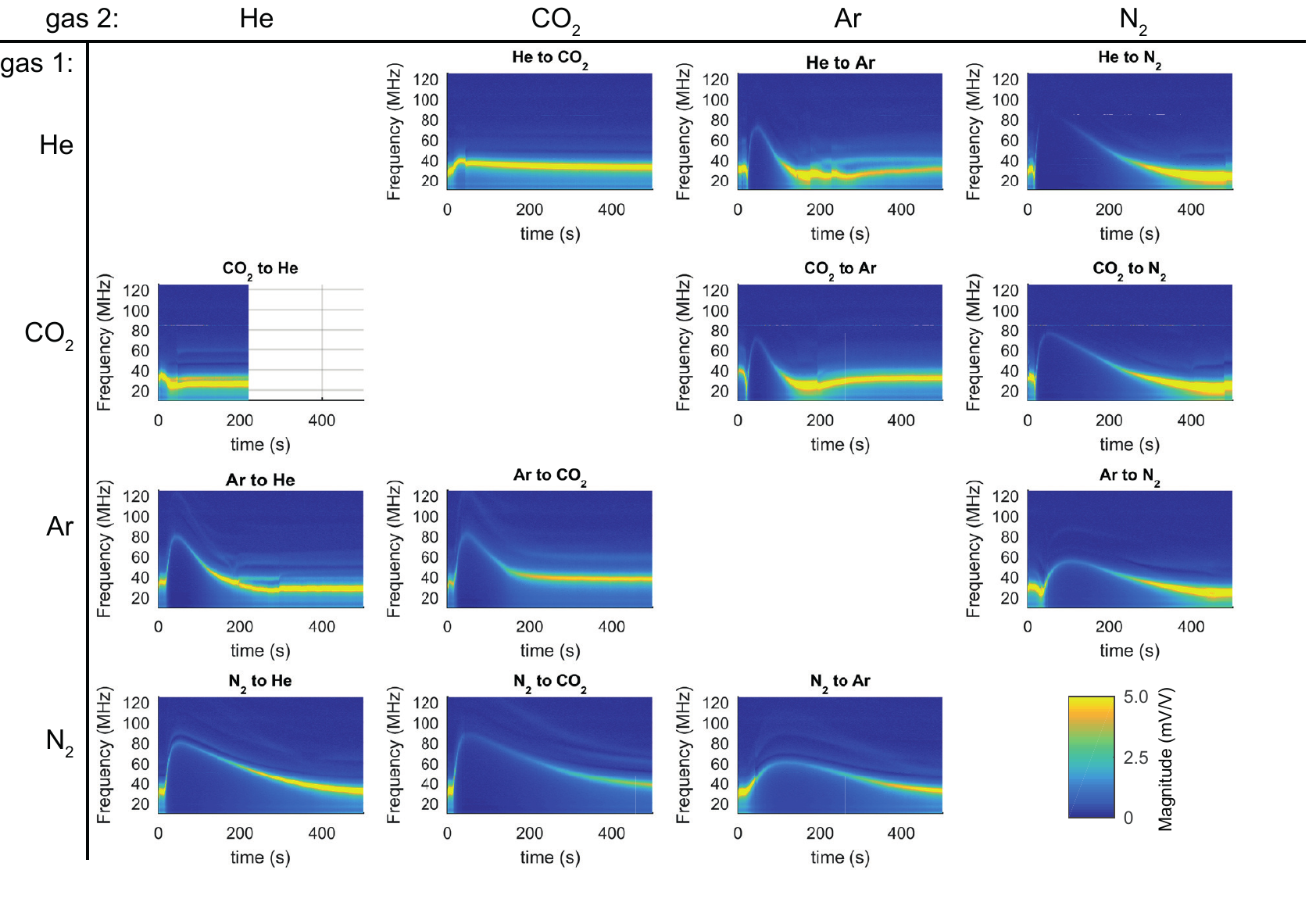}
\caption{Full raw dataset for drum 1 in the main text, measurement was performed at constant 500 mbar chamber pressure. \label{fig:drum1}}
\end{figure*}

\begin{figure*}
\includegraphics{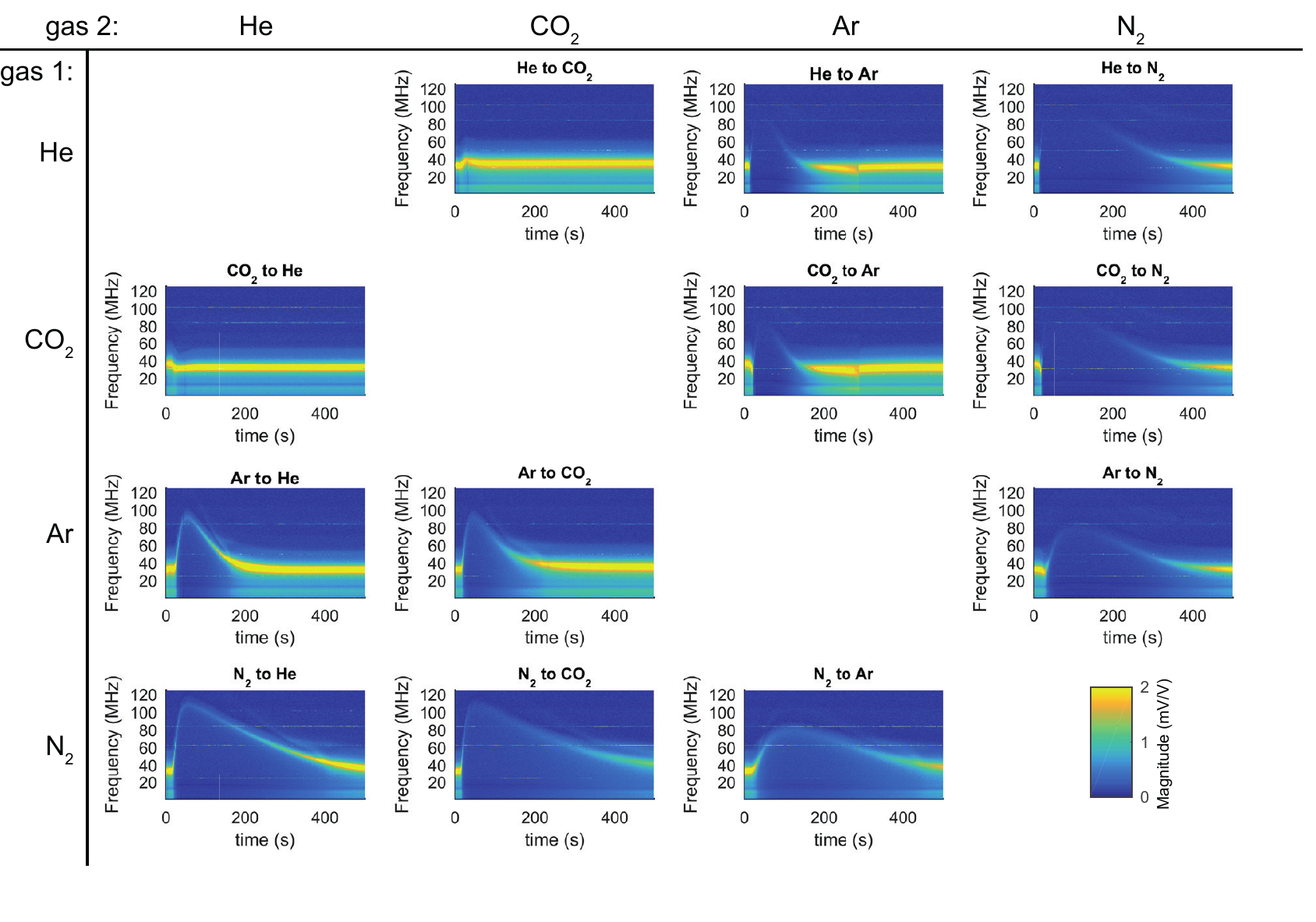}
\caption{Full raw dataset for drum 2 in the main text, measurement was performed at 1000 mbar constant chamber pressure. \label{fig:drum2}}
\end{figure*}

Figure \ref{fig:drum1} shows the full raw dataset for drum 1 in the main text and Fig. \ref{fig:drum2} for drum 2. For both experiments, the figures on the top right show negative pressurer difference while the bottom left figures show positive pressure difference. The nomenclature ''Gas 1 to Gas 2'' means that Gas 1 is initially in the chamber and cavity and the gas in the chamber is replaced by Gas 2. 

\end{document}